\documentclass[10pt,twocolumn]{article}
\usepackage{graphicx} 
\usepackage{amsmath, amssymb}
\usepackage{booktabs}
\usepackage{multirow}
\usepackage{adjustbox}
\usepackage[most]{tcolorbox}
\usepackage{tabularx}
\usepackage{array}
\usepackage{makecell}

\usepackage[hidelinks]{hyperref}

\newtcolorbox{promptschema}{
  breakable,
  colback=white,
  colframe=black,
  boxrule=0.8pt,
  arc=0pt,
  left=4pt,
  right=4pt,
  top=4pt,
  bottom=4pt,
  before skip=8pt,
  after skip=8pt
}

\title{RAG-Match: Retrieval-Augmented Knowledge Injection and Hierarchical Reasoning for Calibrated Semantic Relevance Judgment}

\author{
Hengjun Jiang$^{*,\dagger}$ \and
Liansheng Sun$^{*}$ \and
Yan Jiang$^{*, \ddagger}$ \and
Xiaojie Ke \and
Yongjin Wang \and
Xiangkun Liu \and
Cunxin Gu \and
Jian Xu \and
Guanjun Jiang \\
Qwen Applications Business Group of Alibaba
}

\begin{document}

\maketitle

\begingroup
\renewcommand\thefootnote{}
\footnotetext{{\raggedright $^{*}$ Equal contribution.}}
\footnotetext{{\raggedright $^{\ddagger}$ Work done during an internship at Alibaba.}}
\footnotetext{{\raggedright $^{\dagger}$ Corresponding author: \par \url{jianghengjun.jhj@alibaba-inc.com}.\par}}
\endgroup

\begin{abstract}
Semantic relevance judgment for search is particularly challenging in knowledge-intensive scenarios, where accurate ranking requires not only semantic matching but also background grounding, multi-step reasoning, and well-calibrated decision boundaries. Existing relevance models mainly rely on direct label supervision or shallow semantic similarity, which limits their ability to handle implicit intent, factual equivalence, and fine-grained relevance distinctions. To address this issue, we propose \textsc{RAG-Match}, a three-stage framework that integrates knowledge-augmented pretraining, hierarchical reasoning alignment, and preference-based decision calibration for relevance modeling. The key idea is to first strengthen query-centered semantic grounding, then align the model with structured relevance reasoning, and finally correct decision-level inconsistencies in difficult boundary cases. Experimental results on a real-world search relevance benchmark show that \textsc{RAG-Match} consistently outperforms strong LLM-based baselines across multiple ranking metrics, demonstrating the effectiveness of combining knowledge injection, reasoning supervision, and preference optimization for fine-grained relevance judgment.
\end{abstract}


\section{Introduction}
Relevance judgment is a core problem in information retrieval, especially in search systems where ranking quality directly determines user experience\cite{manning2008introduction}. While traditional retrieval and ranking methods have achieved strong performance on head queries and literal matching scenarios\cite{ponte1998language}, they often struggle when relevance depends on implicit user intent, factual equivalence, or external domain knowledge. For example, a query such as ``2024 Olympics'' may need to be matched with documents referring to ``the 33rd Olympic Games'' even though the lexical overlap is limited\cite{yih2016value}. Recent advances in neural language models have enabled more sophisticated semantic matching \cite{humeau2019poly}, yet integrating explicit knowledge and reasoning capabilities into relevance models remains a significant challenge. In such cases, accurate relevance judgment requires not only semantic matching but also knowledge grounding and reasoning over latent constraints.

Early relevance models mainly relied on lexical overlap and statistical scoring functions such as TF-IDF and BM25\cite{bm25, tfidf}. Later, neural ranking models and pretrained language models substantially improved semantic matching by learning richer interactions between queries and documents\cite{bert, relevance_bert}. However, these approaches still face clear limitations in knowledge-intensive search scenarios. First, many difficult relevance decisions depend on background knowledge that is not explicitly present in the query-document pair. Second, direct label supervision often encourages shortcut learning, where models over-rely on surface-level similarity rather than reasoning about whether the document truly satisfies the query intent. Third, although large language models (LLMs) exhibit strong reasoning ability\cite{llm_ranking, adaptation_search}, naively applying them to relevance judgment often leads to unstable intermediate reasoning, hallucinated constraints\cite{cot_error, hallucination_llm}, and overestimation of documents with high keyword overlap\cite{optimistic_bias}.

A key challenge is that retrieved evidence is often fragmented, whereas relevance judgment requires a compact and query-focused understanding of the underlying knowledge. Simply exposing the model to retrieved passages is insufficient. Instead, the model should learn to map a query to a synthesized supporting document that organizes the most relevant evidence into a coherent form. Moreover, this grounded semantic understanding should be further translated into faithful document-level reasoning and calibrated against hard-boundary errors.

A straightforward alternative would be to adopt RAG with prompting and attach retrieved evidence at inference time. However, this approach remains limited by context window length, inference latency, and the transient nature of in-context knowledge use. In production ranking systems, repeatedly concatenating large amounts of external evidence is often infeasible, and in our deployment setting we do not request a RAG-synthesized document during online inference. This motivates our Phase~I design: instead of relying on external knowledge only at prediction time, we use retrieval-enhanced synthesized documents as a training signal so that query-relevant knowledge can be internalized into the model parameters. In this way, the model acquires a stronger query-centered semantic prior that supports more efficient and accurate downstream relevance judgment.

Motivated by these observations, we propose \textsc{RAG-Match}, a multi-stage framework for semantic relevance judgment that combines retrieval-augmented pretraining, hierarchical reasoning alignment, and preference-based calibration. \textsc{RAG-Match} consists of three phases. In the first phase, we introduce a query-to-synthesized-document pretraining objective. For each query, a RAG system first retrieves the top-$k$ most relevant evidence documents and then synthesizes a document that best answers the query. By training the model to generate this RAG-synthesized document from the query, we inject relevant background knowledge into the model and enable it to learn the mapping from queries to their most informative supporting content. In the second phase, we propose a Hierarchical Reasoning Alignment (HRA) framework that decomposes relevance judgment into two stages. The model first performs query grounding using retrieved knowledge and constructs a structured semantic representation of the query intent and constraints. It then evaluates the candidate document under this grounded frame through multi-dimensional relevance analysis. In the third phase, we introduce Discrepancy-Guided Preference Optimization (DGPO), which uses a small amount of human-annotated data to calibrate hard-boundary decisions and reduce the systematic overestimation tendency of LLM-based relevance predictors. 

Compared with prior relevance models, \textsc{RAG-Match} offers three advantages. First, it incorporates external knowledge into the model parameters through scalable retrieval-augmented supervision, rather than relying solely on runtime retrieval or label-only supervision. Second, it transforms relevance judgment from a direct classification problem into a structured reasoning process, improving both prediction quality and interpretability. Third, it achieves effective calibration with limited human annotations by focusing preference optimization on cases that are most easily confused.

Our main contributions are summarized as follows:

\begin{itemize}
    \item We propose \textsc{RAG-Match}, a three-stage framework for knowledge-intensive relevance modeling that progressively integrates query-centered grounding, hierarchical reasoning alignment, and decision calibration. The framework is designed to address relevance judgment scenarios where accurate ranking requires not only semantic matching, but also background knowledge, multi-step reasoning, and fine-grained boundary discrimination.

    \item We introduce a knowledge-augmented pretraining stage that improves downstream relevance modeling by learning query-to-synthesized-document generation. Rather than serving as a standalone ranking objective, this stage provides stronger semantic grounding and knowledge-aware initialization for subsequent reasoning alignment, leading to improved ranking quality after Phase~II.

    \item We design a hierarchical reasoning alignment stage and a discrepancy-guided preference optimization stage to improve relevance decision quality from complementary perspectives. Phase~II distills structured relevance reasoning from a strong teacher model, while Phase~III further calibrates final relevance judgments by correcting the systematic overestimation tendency observed after supervised reasoning alignment.

    \item We conduct extensive experiments and analysis on a real-world search relevance benchmark. The results show that \textsc{RAG-Match} consistently outperforms strong LLM-based baselines across multiple ranking metrics. Additional analysis further clarifies the distinct roles of the three stages, including the contribution of Phase~I to downstream alignment, the high reliability of teacher-generated CoT supervision, and the effectiveness of DGPO in improving decision calibration.
\end{itemize}

Extensive experiments on a search relevance benchmark demonstrate that \textsc{RAG-Match} consistently improves ranking quality and ranking consistency over strong LLM baselines. Ablation studies further show that the three phases are complementary, with the largest gains coming from hierarchical reasoning alignment and additional improvements from retrieval-augmented pretraining and preference-based calibration.

\section{Related Work}
This work lies at the intersection of relevance modeling in information retrieval, large language models for relevance judgment, retrieval-augmented reasoning, and preference-based model alignment. Below, we review the most relevant lines of research and position \textsc{RAG-Match} among them.

\subsection{Relevance Modeling in Information Retrieval}

Relevance modeling has long been a central problem in information retrieval (IR). Early approaches mainly relied on lexical matching and statistical scoring functions such as TF-IDF and BM25 \cite{bm25, tfidf, robertson1995okapi, robertson2009probabilistic}. These methods are effective for exact matching and head queries, but they often struggle when relevance depends on paraphrasing, implicit intent, or factual equivalence.

To overcome the limitations of lexical matching, neural retrieval and ranking models were introduced to learn distributed semantic representations of queries and documents. Representative early models include DSSM \cite{huang2013learning} and CDSSM \cite{shen2014latent}, which project query-document pairs into dense vector spaces for semantic matching. Later work explored richer interaction architectures, including DRMM \cite{guo2016deep}, MatchPyramid \cite{pang2016text}, and attention-based matching models \cite{yang2016stacked, chen2016thorough}, which improved fine-grained interaction modeling between queries and documents.

The emergence of pretrained language models further advanced neural relevance modeling. BERT-based rerankers, such as BERT-Ranker \cite{nogueira2019passage} and BERT-PLI \cite{dai2019deeper}, demonstrated that deep contextual representations can significantly improve ranking quality. Subsequent work also proposed more efficient architectures for large-scale retrieval and reranking, including ColBERT \cite{khattab2020colbert} and SPLADE \cite{formal2021splade}. Despite these advances, most relevance models are still trained with direct label supervision and therefore remain limited in knowledge-intensive settings where the query-document pair alone does not provide sufficient evidence for accurate judgment.

In contrast to prior relevance models that primarily focus on semantic matching, \textsc{RAG-Match} emphasizes query-centered knowledge grounding and structured reasoning. Our framework uses retrieval-augmented pretraining to inject external knowledge into the model and then performs relevance judgment through hierarchical reasoning rather than direct label prediction alone.

\subsection{Large Language Models for Relevance Judgment}

Large language models (LLMs) have recently emerged as powerful tools for ranking and relevance judgment. Models such as GPT-3 \cite{brown2020language}, T5 \cite{raffel2020exploring}, and later instruction-tuned variants have shown strong zero-shot and few-shot capabilities, making them attractive for low-resource relevance tasks \cite{liu2021pretrain, gao2021making, wei2022flan}. In the retrieval community, works such as RankGPT \cite{sun2023chatgpt} demonstrated that LLMs can perform competitive ranking through pairwise or listwise prompting, while instruction-based retrieval approaches such as TART \cite{asai2023task} showed that task-aware prompting can dynamically adjust matching behavior.

To reduce deployment costs, several studies have explored fine-tuning or distilling smaller LLMs for ranking and reranking. RankLLaMA \cite{ma2023fine} showed that decoder-only LLMs can be adapted to retrieval tasks through supervised fine-tuning, while other work distilled ranking knowledge from stronger teacher models into more efficient students \cite{choi2024rradistilldistillingllmspassage, zhu2023large}. These studies indicate that LLMs provide a promising foundation for relevance modeling beyond traditional discriminative architectures.

However, applying LLMs directly to relevance judgment also introduces new challenges. Their predictions can be sensitive to prompting style, their intermediate reasoning may drift without sufficient grounding, and they often overestimate documents with strong lexical overlap even when key semantic constraints are violated. \textsc{RAG-Match} addresses these issues by combining LLM-based reasoning with retrieval-augmented pretraining and targeted preference alignment, thereby improving both the grounding and calibration of relevance decisions. 

\subsection{Retrieval-Augmented Reasoning\\and Chain-of-Thought Supervisi\-on}

Retrieval-Augmented Generation (RAG) has become a standard paradigm for incorporating external knowledge into language models \cite{gao2023retrieval}. By retrieving evidence from an external corpus and conditioning generation on that evidence, RAG improves factuality and helps language models handle knowledge-intensive tasks. Most prior work has focused on question answering, open-domain generation, or factual reasoning, where retrieved passages are used as context at inference time. In contrast, our work uses retrieval not only as runtime evidence, but also as a source of weak supervision for pretraining.

Chain-of-Thought (CoT) prompting and supervision have further expanded the reasoning capabilities of LLMs \cite{wei2022chain}. By eliciting intermediate reasoning steps before producing a final answer, CoT often improves performance on tasks requiring multi-step inference. Recent studies have explored CoT-style reasoning in ranking and retrieval settings, prompting models to generate rationales before outputting relevance labels \cite{zhuang2023beyond}. Other work has shown that structured or knowledge-driven CoT can improve faithfulness by grounding intermediate reasoning in external evidence.

Despite these advances, most existing approaches either use retrieved evidence as auxiliary context or use CoT as a generic prompting strategy. \textsc{RAG-Match} differs in two important respects. First, in the pretraining stage, we use retrieved evidence documents to construct a RAG-synthesized supervision target and train the model with a query-to-synthesized-document objective, thereby injecting query-centered background knowledge into the model parameters. Second, in the reasoning stage, we organize CoT supervision hierarchically: the model first grounds the query with retrieved evidence and then performs multi-dimensional document analysis under that grounded frame. This explicit decomposition is designed for relevance judgment, where query intent, latent constraints, and partial relevance must be jointly considered.

\subsection{Multi-Stage Training and Align\-ment for Generative Relevance\\Modeling}

Recent work has moved beyond zero-shot LLM ranking and begun to optimize large models as dedicated relevance engines through multi-stage training and post-hoc alignment. This line of research is particularly relevant to industrial search, where relevance judgment often involves hard cases, implicit attributes, and ambiguous decision boundaries.

LREF \cite{Tang2025LREF} is one of the representative frameworks in this direction. It combines high-quality data selection, multi-perspective CoT tuning, and DPO-based de-biasing for e-commerce relevance prediction. LREF shows that reasoning supervision and post-training alignment can substantially improve LLM-based relevance prediction. However, its emphasis is primarily on improving reasoning trajectories and reducing bias after supervised adaptation. By contrast, \textsc{RAG-Match} introduces an additional retrieval-augmented pretraining phase before reasoning supervision. In particular, our Phase~I trains the model to generate a RAG-synthesized document that best answers the query, allowing the model to internalize query-centered background knowledge before downstream reasoning alignment. This makes \textsc{RAG-Match} explicitly knowledge-grounded at the pretraining level, rather than relying only on reasoning supervision at later stages.

ADORE \cite{Fang2025ADORE} further extends multi-stage relevance modeling by combining rule-aware relevance discrimination, error-type-aware hard sample synthesis, and key-attribute-enhanced knowledge distillation. Its design highlights the importance of domain-specific hard cases and explicit attribute grounding. Compared with ADORE, \textsc{RAG-Match} is less centered on handcrafted rule modeling and synthetic hard-sample construction, and instead focuses on two complementary mechanisms: retrieval-augmented knowledge injection through query-to-synthesized-document pretraining, and hierarchical reasoning decomposition through query grounding followed by multi-dimensional document analysis. In other words, while ADORE emphasizes rule-aware discrimination and hard-case synthesis, \textsc{RAG-Match} emphasizes compact knowledge grounding and structured reasoning under a query-centered semantic frame.

LORE \cite{Lu2025LORE} provides a more systematic perspective by arguing that relevance should be decomposed into several core capabilities, including knowledge and reasoning, multimodal matching, and rule adherence. It accordingly proposes a capability-oriented training and evaluation framework. This view is highly aligned with our motivation that relevance judgment is not a single-step matching problem. However, \textsc{RAG-Match} differs from LORE in two key aspects. First, our framework operationalizes knowledge capability through an explicit retrieval-augmented pretraining objective, rather than treating knowledge as one capability dimension in a broader taxonomy. Second, our reasoning process is explicitly hierarchical: the model first constructs a grounded semantic understanding of the query from retrieved evidence and then evaluates the candidate document under that semantic frame. This query-first reasoning protocol is central to our design and is more tightly coupled to graded relevance judgment.

TaoSR1 \cite{Dong2026TaoSR1} pushes this line of work toward online deployment by combining CoT-based supervised fine-tuning, pass@N-based preference optimization, and GRPO-based refinement to improve both reasoning quality and final relevance decisions. Similarly, reinforcement-learning-based generative relevance modeling has also been explored in open-domain and industrial search settings, where relevance assessment is optimized as a multi-step reasoning process with process-aware supervision \cite{Zeng2026XiaohongshuRL}. These approaches demonstrate the value of post-training optimization for relevance alignment. Compared with them, \textsc{RAG-Match} places greater emphasis on the role of retrieval-augmented semantic grounding before post-training alignment. Our framework does not start from reasoning optimization alone; instead, it first builds a compact query-centered knowledge prior through RAG-synthesized document generation, and then uses hierarchical reasoning supervision and discrepancy-guided preference optimization to refine the final relevance boundary.

Overall, prior work has explored reasoning distillation, rule-aware relevance modeling, capability-oriented training, and preference-based calibration for LLM-based ranking. \textsc{RAG-Match} shares the multi-stage spirit of these methods, but differs in its combination of three design choices: (1) retrieval-augmented pretraining through query-to-synthesized-document generation, (2) hierarchical query-first reasoning alignment, and (3) discrepancy-guided calibration on adjacent label confusions. We believe this combination is particularly well suited for semantic relevance judgment in knowledge-intensive search, where external evidence, structured reasoning, and hard-boundary calibration are all essential.

\section{Method}
\begin{figure*}[t]
    \centering
    \includegraphics[width=1.0\linewidth]{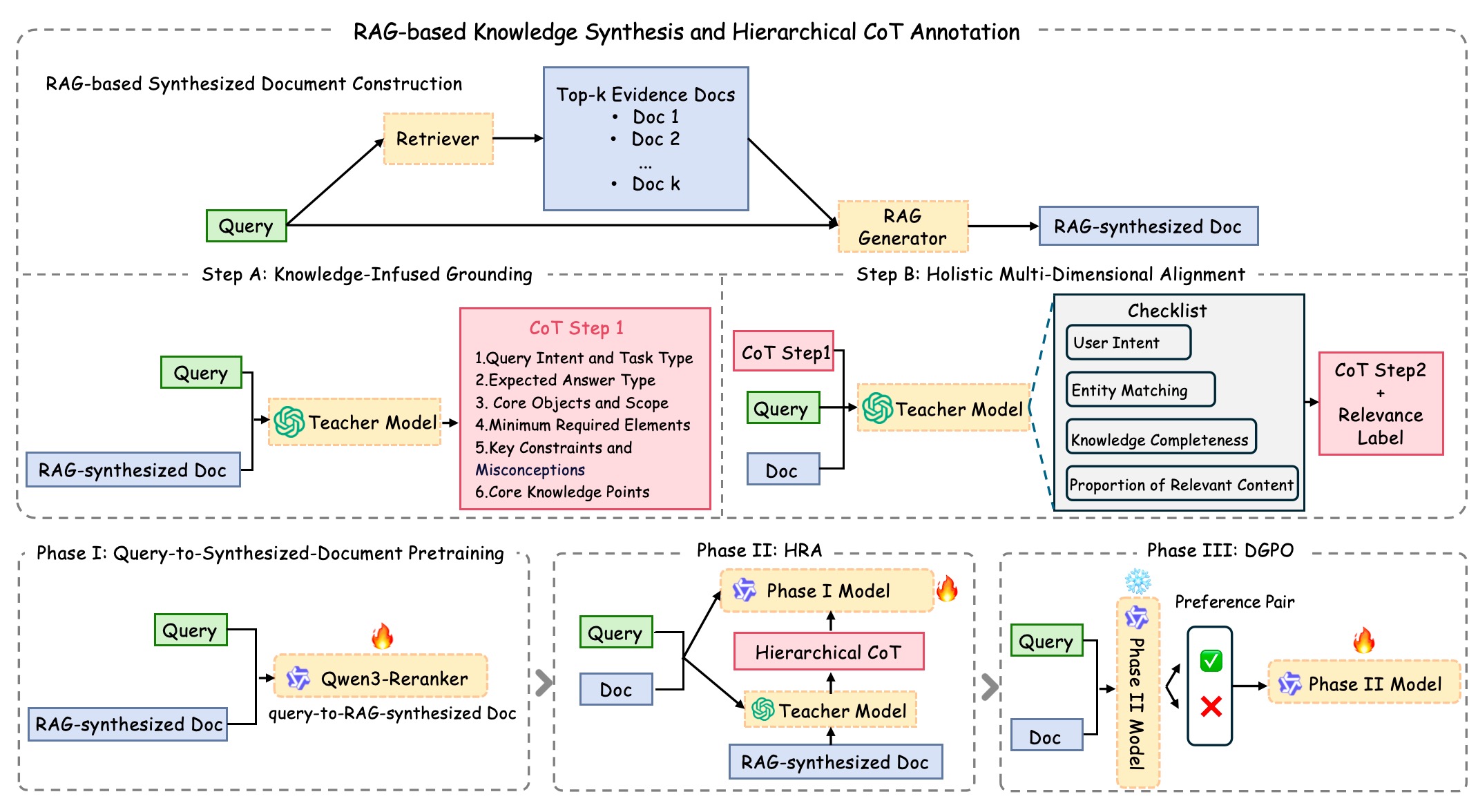}
    \caption{Overview of \textsc{RAG-Match}. For each query, a RAG system retrieves top-$k$ evidence documents and synthesizes a document for Phase~I pretraining. In Phase~II, GPT-5.2 generates hierarchical reasoning supervision through KIG and HMDA. In Phase~III, preference pairs generated by the Phase~II model are used for discrepancy-guided preference optimization.}
    \label{fig:framework}
\end{figure*}
In this section, we present \textsc{RAG-Match}, a multi-stage framework for semantic relevance judgment in knowledge-intensive search. The framework consists of three phases: (1) retrieval-augmented pretraining for query-centered semantic grounding, (2) hierarchical reasoning alignment for structured relevance prediction, and (3) discrepancy-guided preference optimization for hard-boundary calibration. Figure~\ref{fig:framework} illustrates the overall pipeline, including RAG-based synthesized document construction, hierarchical reasoning annotation, and the three-stage training process.

\subsection{Problem Definition}

We study query-document relevance judgment with retrieval-augmented evidence. Given a query $q \in \mathcal{Q}$, a candidate document $d \in \mathcal{D}$, and a set of retrieved evidence documents $\mathcal{E} = \{e_1, \dots, e_m\}$ obtained from a knowledge base $\mathcal{B}$, the goal is to predict an ordinal relevance label $y \in \{0,1,2,3\}$, where $3$, $2$, $1$, and $0$ denote \textit{Excellent}, \textit{Good}, \textit{Fair}, and \textit{Irrelevant}, respectively.

Rather than treating relevance prediction as a direct classification problem, \textsc{RAG-Match} formulates it as a structured generative reasoning process. Specifically, the model first generates a query-grounding rationale $c_1$ conditioned on the retrieved evidence, and then generates a document-level rationale $c_2$ together with the final label $y$ conditioned on the candidate document and the grounding rationale. The joint probability is factorized as
\begin{equation}
p_\theta(c_1, c_2, y \mid q, d, \mathcal{E})
=
p_\theta(c_1 \mid q, \mathcal{E})
\cdot
p_\theta(c_2, y \mid q, d, c_1).
\end{equation}

During inference, the relevance score is derived from the generated label $y$, while $c_1$ and $c_2$ serve as interpretable rationales for the final decision.

\subsection{Retrieval-Augmented Knowledge Construction}

Retrieved knowledge plays two roles in \textsc{RAG-Match}: it provides weak supervision for pretraining and serves as grounding evidence during reasoning. To obtain such knowledge, we build a retrieval module over a domain knowledge base $\mathcal{B}$, which consists of structured or semi-structured text fragments extracted from the search corpus and auxiliary resources.

For each query $q$, the retriever returns a set of top-$k$ evidence documents:
\begin{equation}
\mathcal{E} = \mathrm{Top}\text{-}k(R(q, \mathcal{B})),
\end{equation}
where $R$ denotes the retrieval function. In our implementation, the retriever is a bi-encoder trained with contrastive learning, and approximate nearest neighbor search(e.g., FAISS) is used for efficient retrieval.

The retrieved evidence documents are not treated as final supervision targets by themselves. Instead, they are first used by a RAG generator to synthesize a compact query-focused document. This RAG-synthesized document then serves different roles in the subsequent stages: in Phase~I, it is used as the supervision target for query-to-document pretraining, while in Phase~II, it is further used in KIG to support query grounding before document-level relevance analysis.

\subsection{Knowledge-Augmented Semantic\\Grounding via Query-to-Synthe\-sized-Document Pretraining}

Literal query-document matching is often insufficient for complex or underspecified search queries, especially when the query intent depends on dispersed background knowledge rather than surface lexical overlap. To address this issue, we introduce a knowledge-augmented pretraining stage based on query-to-synthesized-document generation. Instead of directly predicting relevance labels, this stage trains the model to map a query to a compact document synthesized by a RAG system from top-$k$ retrieved evidence documents. In this way, the model learns to internalize query-centered background knowledge and acquires grounded semantic priors before entering downstream relevance reasoning.

\subsubsection{RAG-Synthesized Pretraining Corpus Construction}

We construct a large-scale pretraining corpus $\mathcal{D}_{pre}$ from 10 million search queries sampled from historical logs. Since these queries do not have manual relevance annotations, we derive weak supervision through a retrieval-augmented generation pipeline.

For each query $q_i$, we first use a retriever to obtain the top-$k$ most relevant evidence documents from the knowledge base $\mathcal{B}$:
\begin{equation}
\mathcal{E}_i = \mathrm{Top}\text{-}k(R(q_i, \mathcal{B})).
\end{equation}

We then feed the query $q_i$ together with the retrieved evidence set $\mathcal{E}_i$ into a RAG generator, which synthesizes a single document $s_i$ that best answers the query:
\begin{equation}
s_i = G_{\text{RAG}}(q_i, \mathcal{E}_i).
\end{equation}

The resulting synthesized document $s_i$ is not necessarily identical to any individual retrieved document. Instead, it provides a compact, query-focused, and knowledge-grounded summary of the evidence most relevant to the query. We therefore use the pair $(q_i, s_i)$ as a weakly supervised pretraining instance.

\subsubsection{Query-to-Synthesized-Document Pretraining Objective}

To inject query-centered background knowledge into the model, we design a query-to-synthesized-document objective. Given a query $q_i$, the model is trained to generate the corresponding RAG-synthesized document $s_i$:
\begin{equation}
p_\theta(s_i \mid q_i)
=
\prod_{t=1}^{|s_i|}
p_\theta(s_i^t \mid s_i^{<t}, q_i).
\end{equation}

The pretraining loss is defined as
\begin{equation}
\mathcal{L}_{pre}(\theta)
=
-
\mathbb{E}_{(q,s)\in\mathcal{D}_{pre}}
\left[
\sum_{t=1}^{|s|}
\log p_\theta(s^t \mid s^{<t}, q)
\right].
\end{equation}

This objective encourages the model to learn the latent mapping from a query to its most informative supporting document, rather than merely memorizing lexical correlations. Since each synthesized target is constructed from top-$k$ retrieved evidence documents, the model is exposed to query-relevant entities, attributes, and factual relations in a compact form. 

Phase~I is not designed to directly optimize the final ranking metrics. Instead, its role is to improve the model initialization by reshaping the intermediate semantic representation used in downstream relevance modeling. The query-to-synthesized-document objective encourages the model to construct a query-centered information state that captures implicit intent, relevant background knowledge, and the semantic attributes that a relevant document is expected to satisfy. Compared with direct relevance labels, this generative objective provides denser and more structured supervision, because it teaches the model not only whether a document is relevant, but also what information a query should evoke.

From this perspective, Phase~I can be interpreted as learning a stronger query representation before discriminative alignment. Let $z_q=f_\theta(q)$ denote the internal query representation. By requiring $z_q$ to support the generation of a query-focused synthesized document, the model is encouraged to preserve query intent, background facts, and semantically salient content structure in $z_q$. This provides a stronger basis for downstream relevance prediction $y=g(z_q,d)$ than direct label supervision alone. In other words, Phase~I helps discrimination not by directly learning the final ranking boundary, but by learning a query-conditioned semantic scaffold that makes later relevance reasoning easier and more stable.

From an optimization perspective, Phase~I also serves as a task-relevant warm start. Without this stage, the model in Phase~II must simultaneously infer latent query requirements, organize supporting knowledge, and learn decision-oriented reasoning from relatively sparse supervision. By pretraining the model to first recover a query-centered semantic support space, Phase~I decouples semantic grounding from later reasoning alignment, thereby making the downstream optimization problem easier. This also explains why the benefit of Phase~I is mainly reflected in improved NDCG after Phase~II, while its additional effect on nPNR is limited: Phase~I primarily strengthens semantic grounding and graded ranking quality, whereas pairwise ranking consistency is more directly shaped by the explicit hierarchical supervision introduced in Phase~II.

\subsection{Hierarchical Reasoning Alignme\-nt}

After retrieval-augmented pretraining, the model has acquired query-centered semantic priors, but it still needs to transform these priors into faithful relevance decisions. Directly reasoning over the full tuple $(q, d, s)$ can be suboptimal, because the model may overfit superficial overlap between the query and the candidate document before establishing a precise understanding of the query intent. To address this problem, we introduce Hierarchical Reasoning Alignment (HRA), which decomposes relevance judgment into two stages: query grounding and document-level alignment.

\subsubsection{Stage A: Knowledge-Infused Grounding}

In the first stage, the model performs Knowledge-Infused Grounding (KIG) using only the query $q$ and the RAG-synthesized document s. The purpose of this stage is to construct a grounded semantic state before the candidate document is introduced. This design reduces reasoning drift\cite{wang2023knowledgedrivencotexploringfaithful} and forces the model to explicitly organize the latent constraints underlying the query.

We instantiate the grounding process using a teacher LLM, which maps $(q, s)$ to a structured six-field representation:
\begin{equation}
\mathbf{G} = [I, T, O, E, C, S],
\end{equation}
where $I$ denotes the query intent, $T$ denotes the expected answer type, $O$ denotes the core object range, $E$ denotes the minimal answer elements, $C$ denotes critical constraints, and $S$ denotes a compact synopsis of the retrieved knowledge. Formally,
\begin{equation}
\mathbf{G} = f_{\text{teach}}^{(1)}(q, s).
\end{equation}

This representation serves as an explicit semantic frame for the next stage. It summarizes what the user is asking for, what constraints must be satisfied, and what background knowledge is relevant to the judgment.

\subsubsection{Stage B: Holistic Multi-Dimensional Alignment}

In the second stage, the model evaluates the candidate document $d$ under the grounded semantic frame $\mathbf{G}$. Instead of making a direct holistic judgment, we decompose relevance into four complementary dimensions:
\begin{itemize}
    \item \textbf{Intent Consistency} ($\Phi_{IC}$): whether the document satisfies the expected answer type and core intent of the query;
    \item \textbf{Entity Fidelity} ($\Phi_{EF}$): whether the entities and concepts in the document match the grounded object range, including aliases and hierarchical relations;
    \item \textbf{Logical Completeness} ($\Phi_{LC}$): whether the document covers the minimal answer elements required by the query;
    \item \textbf{Information Density} ($\Phi_{ID}$): whether the document contains concentrated relevant evidence rather than noisy or misleading content.
\end{itemize}

We use a teacher LLM to generate a structured alignment rationale
\begin{equation}
\mathbf{A} = f_{\text{teach}}^{(2)}(q, d, \mathbf{G}),
\end{equation}
which summarizes the document analysis across these dimensions and supports the final relevance label.

This decomposition is designed to capture common failure modes in relevance judgment, including superficial topic overlap, partial answer coverage, entity mismatch, and noisy content. By enforcing dimension-wise analysis, HRA improves both decision quality and interpretability.

\subsubsection{Reasoning Distillation via Supervised Fine-Tuning}

To transfer the HRA reasoning protocol to the student model, we build a reasoning-annotated dataset $\mathcal{D}_{cot}$ using the teacher LLM. Each instance contains a query $q$, a candidate document $d$, a grounding rationale $\mathbf{G}$, a document-level alignment rationale $\mathbf{A}$, and the final relevance label $y$.

The student model is then fine-tuned to generate the full reasoning target $t = (\mathbf{G}, \mathbf{A}, y)$ conditioned on the input $x = (q, d)$:
\begin{equation}
\mathcal{L}_{sft}(\theta)
=
-
\mathbb{E}_{(x,t)\in\mathcal{D}_{cot}}
\log p_\theta(t \mid x).
\end{equation}

By distilling structured reasoning trajectories rather than labels alone, HRA teaches the model to perform relevance judgment through explicit query grounding and document-level analytical reasoning.

\subsection{Discrepancy-Guided Preference\\Optimization}
Although HRA equips the model with hierarchical reasoning supervision, it does not fully resolve decision-level calibration errors in final relevance prediction. In particular, we observe that the Phase~II model tends to overestimate relevance on ambiguous or boundary cases, resulting in an upward scoring tendency in its final labels. To address this issue, we introduce Phase~III, namely discrepancy-guided preference optimization (DGPO)\cite{rafailov2023direct}, which further calibrates the model's decision boundaries by encouraging preferences toward more reliable relevance outcomes. As shown later in our analysis, this stage effectively reduces systematic overestimation and improves the robustness of final relevance judgments.

\subsubsection{Preference Pair Construction}

We construct preference pairs from a small set of human-annotated samples. For each sample $(q, d, y^\ast)$, where $y^\ast$ is the gold relevance label, we treat $y^\ast$ as the preferred label $y_w$ and sample an adjacent incorrect label $y_l$ from the valid neighborhood of $y^\ast$. This design focuses the preference learning process on hard confusions rather than obviously incorrect alternatives.

Using the Phase-II model, we generate two reasoning trajectories conditioned on these labels:
\begin{itemize}
    \item a preferred trajectory $(y_w, r_w)$, where $r_w$ is the rationale consistent with the gold label;
    \item a dispreferred trajectory $(y_l, r_l)$, where $r_l$ corresponds to a plausible but incorrect judgment.
\end{itemize}

The resulting dataset
\begin{equation}
\mathcal{D}_{pair} = \{(q,d,y_w,r_w,y_l,r_l)\}
\end{equation}
contains preference pairs centered on difficult decision boundaries.

\subsubsection{Preference-Based Calibration}

We optimize the student policy $\pi_\theta$ against a reference policy $\pi_{ref}$. For a reasoning trajectory $(y, r)$, we define
\begin{equation}
h(y,r)
=
\log \frac{\pi_\theta(y,r \mid q,d)}{\pi_{ref}(y,r \mid q,d)}.
\end{equation}

The DPO objective is
\begin{equation}
\mathcal{L}_{dpo}(\theta)
=
-
\mathbb{E}_{\mathcal{D}_{pair}}
\left[
\log \sigma \left(
\beta h(y_w,r_w) - \beta h(y_l,r_l)
\right)
\right],
\end{equation}
where $\sigma$ is the sigmoid function and $\beta$ controls the alignment strength.

Because online deployment primarily depends on the final relevance label, we further add an auxiliary label prediction term:
\begin{equation}
\mathcal{L}_{dpo+}(\theta)
=
\mathcal{L}_{dpo}
-
\mathbb{E}_{(q,d,y_w)\in\mathcal{D}_{pair}}
\log \pi_\theta(y_w \mid q,d).
\end{equation}

This final calibration stage encourages the model to prefer more faithful reasoning trajectories and sharper decision boundaries, thereby reducing the systematic overestimation tendency on ambiguous relevance cases.

\section{Experiments}

In this section, we evaluate the performance of \textsc{RAG-Match} through comparative benchmarks and ablation studies. We show that the proposed combination of retrieval-augmented pretraining, hierarchical reasoning alignment, and discrepancy-guided preference optimization substantially improves both ranking quality and ranking consistency in knowledge-intensive relevance judgment.

\subsection{Datasets and Metrics}

For the pretraining stage, we sampled 10 million search queries from historical logs. For each query, we first retrieved the top-$k$ most relevant evidence documents using a retriever and then used a RAG generator to synthesize a document that best answers the query. The resulting query-document pairs $(q, s)$ were used to train the query-to-synthesized-document objective in Phase~I.

For the training set of the Hierarchical Reasoning Alignment task, we used a strong teacher LLM to annotate and reason over 25{,}102 unlabeled samples. Each sample is a triplet $\langle q, d, s \rangle$ consisting of a query $q$, a candidate document $d$, and a RAG-synthesized document s. Given this triplet, the teacher model produces the structured grounding rationale $\mathbf{G}$, the document-level alignment rationale $\mathbf{A}$, and the final relevance label $y$. 

We use a small set of 3{,}000 human-annotated samples to train the Discrepancy-Guided Preference Optimization task. For each sample, preference pairs are constructed around adjacent label confusions, and the corresponding reasoning trajectories are generated by the Phase~II model.

The label distribution is detailed in Table~\ref{tab:label_distribution}, showing a representative mix of typical search scenarios.

\begin{table*}[t]
    \centering
    \caption{Label distribution of the training and test sets. The relevance labels are defined as 3 = Excellent, 2 = Good, 1 = Fair, and 0 = Irrelevant.}
    \label{tab:label_distribution}
    \begin{tabular}{lcccc}
        \toprule
        \textbf{Split} & \textbf{Excellent (3)} & \textbf{Good (2)} & \textbf{Fair (1)} & \textbf{Irrelevant (0)} \\
        \midrule
        HRA Training Set & 27.09\% & 34.57\% & 24.20\% & 14.14\% \\
        DGPO Training Set & 18.76\% & 37.84\% & 22.74\% & 20.67\% \\
        Test Set & 13.60\% & 32.18\% & 23.67\% & 30.56\% \\
        \bottomrule
    \end{tabular}
\end{table*}

To evaluate the model’s generalization in real-world scenarios, we construct a manually annotated test set of 1,728 query-document pairs collected from search logs. The test set covers 300 queries and is designed to include both randomly sampled queries and medium-/long-tail search queries, so as to reflect realistic and challenging search conditions, including ambiguous user intents. We note that the test set size is limited by the high cost and relatively low efficiency of professional manual annotation, especially for fine-grained relevance judgment. Therefore, we focus on building a carefully curated evaluation set with high annotation quality rather than a larger but noisier benchmark. All query-document pairs in the test set were annotated by professional judges to provide a gold standard for evaluation.

We adopt two categories of evaluation metrics:
\begin{itemize}
    \item \textbf{Ranking Quality:} We report NDCG@K ($K \in \{1,3,5,10\}$) to assess the quality of top-ranked results \cite{wang2013theoretical}.
    \item \textbf{Ranking Consistency (nPNR):} Following prior work, we also consider pairwise ranking consistency as an auxiliary evaluation signal. Instead of directly reporting the raw Positive-Negative Ratio (PNR), which is sensitive to the number of positive and negative document pairs and lacks a standardized range, we report its normalized form:
\begin{equation}
\text{nPNR} = \frac{\text{PNR}}{1+\text{PNR}} = \frac{N_{+}}{N_{+}+N_{-}},
\end{equation}
where $N_{+}$ and $N_{-}$ denote the numbers of correctly and incorrectly ordered document pairs, respectively. This normalization maps the metric into the range $(0,1)$, making it easier to interpret and compare across models. A higher nPNR indicates better pairwise ranking consistency.
\end{itemize}

\subsection{Baselines}

To ensure a rigorous and fair evaluation, we compare \textsc{RAG-Match} against several strong large language model baselines of comparable scale. All baseline models are fine-tuned on the same labeled training data using a standard label-only supervised fine-tuning objective before evaluation.
\begin{itemize}
    \item \textbf{DeepSeek-R1-0528 (8B)} \cite{DeepSeekAI2025DeepSeekR1}: a distilled reasoning model optimized for complex logical inference.
    \item \textbf{GLM-4-9B-0414 (9B)} \cite{glm2024chatglm}: a dense bilingual model with strong semantic understanding and long-context modeling ability.
    \item \textbf{MiniMax-SynLogic (8B)} \cite{liu2025synlogicsynthesizingverifiablereasoning}: a model specialized in structural logic and symbolic reasoning for text matching.
    \item \textbf{Qwen3-8B-Reranker (8B)} \cite{zhang2025qwen3embeddingadvancingtext}: our backbone model trained with label-only supervised fine-tuning, serving as the main base model comparison.
\end{itemize}

\subsection{Implementation Details}

We employ Qwen3-Reranker-8B as the backbone model for \textsc{RAG-Match}. Phase~I (knowledge-augmented pretraining) is conducted on 10M queries. For each query, we retrieve top-$k$ evidence documents and use a RAG generator to synthesize a query-focused supervision document. The backbone model is then trained with the query-to-synthesized-document objective using the AdamW optimizer with a learning rate of $1.5 \times 10^{-5}$.

Phase~II (Hierarchical Reasoning Alignment) uses GPT-5.2 as the teacher LLM to construct structured reasoning trajectories. Specifically, given a triplet $\langle q, d, s \rangle$, GPT-5.2 generates the grounding rationale $\mathbf{G}$, the document-level alignment rationale $\mathbf{A}$, and the final relevance label $y$, which together form the reasoning-annotated dataset $\mathcal{D}_{cot}$. The student model is then fine-tuned on $\mathcal{D}_{cot}$ with a maximum sequence length of 2048. We choose GPT-5.2 as the teacher model in Phase~II because of its strong reasoning capability and high-quality CoT generation. In our manual inspection, GPT-5.2 achieves 93.3\% accuracy on CoT generation, substantially higher than its relevance label accuracy of 76.6\%, suggesting that it is particularly suitable for providing structured reasoning supervision.

Phase~III (Discrepancy-Guided Preference Optimization) is built on top of the Phase~II student model. Specifically, for each human-annotated sample, we construct a preferred label and an adjacent dispreferred label, and then use the Phase~II model to generate the corresponding reasoning trajectories. These preferred--dispreferred trajectory pairs are used for DPO-based calibration with $\beta = 0.1$, allowing the model to further refine subtle relevance boundaries while remaining consistent with the reasoning protocol learned in Phase~II.

Training is performed on an $8 \times$ NVIDIA A800 cluster.

\subsection{Main Results}

Table~\ref{tab:main_results} presents the main comparison results. \textsc{RAG-Match} consistently outperforms all baseline models across all reported NDCG positions and nPNR. In particular, \textsc{RAG-Match} achieves the best performance in both top-rank quality and ranking consistency, indicating that the proposed framework effectively bridges the gap between retrieved knowledge, structured reasoning, and final relevance judgment.

\begin{table*}[t]
    \centering
    \caption{Main results on the search relevance benchmark. All baseline models are fine-tuned on the same labeled training data using label-only supervised fine-tuning. The best results are in bold.}
    \label{tab:main_results}
    \begin{adjustbox}{width=\textwidth}
    \begin{tabular}{lcccccc}
        \toprule
        \textbf{Model} & \textbf{NDCG@1$\uparrow$} & \textbf{NDCG@3$\uparrow$} & \textbf{NDCG@5$\uparrow$} & \textbf{NDCG@10$\uparrow$} & \textbf{nPNR$\uparrow$} \\
        \midrule
        DeepSeek-R1-0528 (8B) & 0.815 & 0.867 & 0.883 & 0.906 & 0.783 \\
        GLM-4-9B-0414 (9B) & 0.846 & 0.877 & 0.891 & 0.915 & 0.801 \\
        MiniMax-SynLogic (8B) & 0.837 & 0.87 & 0.891 & 0.908 & 0.792 \\
        Qwen3-8B-Reranker (8B) & 0.826 & 0.878 & 0.892 & 0.912 & 0.796 \\
        \midrule
        \textsc{RAG-Match(8B)} & \textbf{0.902} & \textbf{0.907} & \textbf{0.916} & \textbf{0.935} & \textbf{0.833} \\
        \bottomrule
    \end{tabular}
    \end{adjustbox}
\end{table*}

Notably, \textsc{RAG-Match} brings clear improvements on both NDCG and nPNR, indicating that the proposed framework improves not only top-ranked retrieval quality but also pairwise ranking consistency. This suggests that the model is better able to capture fine-grained relevance distinctions and maintain more reliable decision boundaries in hard and ambiguous search scenarios.

\subsection{Ablation Study}

Table~\ref{tab:ablation_macro} shows the macro-level ablation results of different components in \textsc{RAG-Match}. Overall, the full model achieves the best performance across all ranking metrics, indicating that the three stages contribute complementary improvements.

Compared with the base model (Row 1), directly appending the RAG-synthesized document at inference time (Row 2) substantially improves all metrics, confirming that retrieved external knowledge is useful for relevance estimation. However, this inference-time RAG baseline still underperforms the label-only SFT model (Row 3) and remains clearly weaker than all multi-stage variants (Rows 4--6). This suggests that simple RAG-style prompting provides only transient in-context evidence and cannot replace task-specific training.

Starting from the label-only SFT baseline, introducing Phase~II (Row 4) brings consistent gains on both NDCG and nPNR, demonstrating the effectiveness of hierarchical reasoning alignment. Adding Phase~I before Phase~II (Row 5) further improves all NDCG metrics, while nPNR remains nearly unchanged, indicating that Phase~I mainly strengthens query-centered semantic grounding and improves overall ranking quality rather than directly enhancing pairwise consistency. Finally, incorporating Phase~III (Row 6) yields the best overall results, showing that discrepancy-guided preference optimization further improves difficult relevance judgments through better decision calibration.

Taken together, these results show that the gain of \textsc{RAG-Match} does not come merely from attaching extra retrieved context at inference time. Instead, it comes from progressively internalizing retrieval-enhanced knowledge, aligning the model with structured relevance reasoning, and calibrating final relevance decisions.

\subsubsection{Effect of Phase~I on Phase~II Alignment}

We do not report standalone results for Phase~I, because its objective is not to directly optimize the final ranking metrics. Instead, Phase~I is designed to provide a knowledge-augmented initialization through the query-to-synthesized-document objective, which is expected to facilitate the subsequent hierarchical reasoning alignment in Phase~II. Therefore, we assess its contribution by comparing a Phase~II-only model with a model trained using Phase~I followed by Phase~II.

\begin{table*}[t]
    \centering
    \caption{Macro ablation results of \textsc{RAG-Match}. Phase~I denotes query-to-synthesized-document pretraining, Phase~II denotes Hierarchical Reasoning Alignment, and Phase~III denotes Discrepancy-Guided Preference Optimization. For fairness, we additionally evaluate an inference-time RAG baseline that appends the RAG-synthesized document during prediction only, without applying the proposed multi-stage training pipeline.}
    \label{tab:ablation_macro}
    \begin{adjustbox}{width=\textwidth}
    \begin{tabular}{lcccc|cccccc}
        \toprule
        \textbf{Row} & \textbf{Phase I} & \textbf{Phase II} & \textbf{Phase III} & \textbf{Infer-time RAG} & \textbf{NDCG@1$\uparrow$} & \textbf{NDCG@3$\uparrow$} & \textbf{NDCG@5$\uparrow$} & \textbf{NDCG@10$\uparrow$} & \textbf{nPNR$\uparrow$} \\
        \midrule
        1 & $\times$ & $\times$ & $\times$ & $\times$ & 0.668 & 0.744 & 0.784 & 0.827 & 0.618 \\
        2 & $\times$ & $\times$ & $\times$ & \checkmark & 0.764 & 0.842 & 0.866 & 0.892 & 0.736 \\
        3 & $\times$ & Label-only SFT & $\times$ & $\times$ & 0.826 & 0.878 & 0.892 & 0.912 & 0.796 \\
        4 & $\times$ & \checkmark & $\times$ & $\times$ & 0.853 & 0.891 & 0.901 & 0.914 & 0.810 \\
        5 & \checkmark & \checkmark & $\times$ & $\times$ & 0.882 & 0.895 & 0.908 & 0.925 & 0.809 \\
        6 & \checkmark & \checkmark & \checkmark & $\times$ & \textbf{0.902} & \textbf{0.907} & \textbf{0.916} & \textbf{0.935} & \textbf{0.833} \\
        \bottomrule
    \end{tabular}
    \end{adjustbox}
\end{table*}

As shown in Table~\ref{tab:ablation_macro}, incorporating Phase~I leads to further improvements in NDCG after Phase~II training, while the nPNR results remain largely comparable. This suggests that the synthesized-document pretraining stage mainly benefits the model's overall ranking quality by strengthening query-centered semantic grounding and enriching background knowledge before reasoning alignment. A plausible explanation for the relatively stable nPNR is that pairwise ranking consistency is more directly shaped by the explicit hierarchical reasoning supervision introduced in Phase~II. In addition, Phase~I may primarily refine graded relevance estimation and the relative ordering of top-ranked documents, which is more likely to improve NDCG than to substantially change the binary pairwise outcomes reflected by nPNR.

\subsubsection{Analysis of Phase II: Hierarchical Reasoning Alignment}

Phase~II is a major source of performance gain in the overall framework. Compared with the label-only supervised fine-tuning baseline, enabling HRA substantially improves both NDCG and nPNR, showing that structured reasoning alignment plays a critical role in relevance prediction quality. At the same time, the later Phase~III stage brings gains of a comparable scale on NDCG and even larger improvements on nPNR, suggesting that reasoning alignment and decision calibration contribute in complementary ways. These results indicate that explicit reasoning supervision is much more effective than direct label fitting in complex search scenarios, where the model must verify intent satisfaction, entity scope, and answer completeness rather than rely on shallow semantic overlap.

We further analyze the internal structure of Phase~II in Table~\ref{tab:ablation_hra}. Adding Stage~A to Stage~B consistently improves both ranking quality and ranking consistency, indicating that query grounding provides an essential semantic frame for subsequent document analysis. Without Stage~A, the model tends to align directly against the candidate document without first establishing a precise understanding of the query, which increases the risk of reasoning drift and superficial matching.

\begin{table*}[t]
    \centering
    \caption{Ablation study on the internal structure of Phase~II. Stage~A denotes Knowledge-Infused Grounding (KIG), and Stage~B denotes Holistic Multi-Dimensional Alignment (HMDA).}
    \label{tab:ablation_hra}
    \begin{adjustbox}{width=\textwidth}
    \begin{tabular}{lcc|cccccc}
        \toprule
        \textbf{Setting} & \textbf{Stage A} & \textbf{Stage B} & \textbf{NDCG@1$\uparrow$} & \textbf{NDCG@3$\uparrow$} & \textbf{NDCG@5$\uparrow$} & \textbf{NDCG@10$\uparrow$} & \textbf{nPNR$\uparrow$} \\
        \midrule
        Direct reasoning baseline & $\times$ & $\times$ & 0.826 & 0.878 & 0.892 & 0.912 & 0.796 \\
        Stage B only & $\times$ & \checkmark & 0.840 & 0.867 & 0.885 & 0.907 & 0.799 \\
        Stage A + Stage B & \checkmark & \checkmark & \textbf{0.853} & \textbf{0.891} & \textbf{0.901} & \textbf{0.914} & \textbf{0.810} \\
        \bottomrule
    \end{tabular}
    \end{adjustbox}
\end{table*}

\begin{table*}[t]
\centering
\caption{Quantification of the upward scoring tendency for the Phase~II model and the model after Phase~III. Phase~III substantially reduces overestimation errors, as reflected by lower OverScore Rate and Mean Score Bias.}
\label{tab:bias_analysis}
\begin{tabular}{lccc}
\toprule
\textbf{Model} & \textbf{OverScore Rate} & \textbf{UnderScore Rate} & \textbf{Mean Score Bias} \\
\midrule
Phase~I + Phase~II & 0.5023 & 0.0683 & 0.5602 \\
Phase~I + Phase~II + Phase~III & \textbf{0.3356} & 0.1423 & \textbf{0.2378} \\
\bottomrule
\end{tabular}
\end{table*}

\subsubsection{Analysis of Phase III: Discrepancy-\\Guided Preference Optimization}

Phase~III (DGPO) serves as the final calibration layer. Comparing the full framework with its supervised fine-tuning predecessor shows that DGPO further suppresses the systematic overestimation tendency inherent in LLM-based relevance predictors. By maximizing the preference margin between the faithful trajectory and the plausible but incorrect trajectory, DGPO improves consistency on hard-boundary cases and leads to the best overall ranking performance. This confirms that discrepancy-guided preference learning is particularly effective for resolving subtle label confusions, such as \textit{Fair} vs.\ \textit{Good}, which are common in real-world search relevance judgment.

The motivation for introducing DGPO in Phase~III is also supported by our manual analysis of the Phase~II teacher annotations. Specifically, GPT-5.2 achieves high accuracy in generating CoT rationales (93.3\%), but its relevance label accuracy is noticeably lower (76.6\%). Moreover, after Phase~II training, the student model attains a CoT accuracy that is largely comparable to that of the teacher, suggesting that the reasoning patterns are effectively distilled. However, our case analysis reveals that some examples contain correct CoT explanations but incorrect final labels. This indicates that even high-quality reasoning supervision does not fully eliminate label noise or decision-boundary errors. DGPO is therefore introduced to further correct such discrepancies by optimizing preference signals over competing reasoning outcomes, thereby improving the robustness of final relevance decisions.

Beyond the gains on the main ranking metrics, Phase~III also improves the calibration of final relevance decisions. We further quantify whether the Phase~II model exhibits an upward scoring tendency in final relevance prediction. Specifically, given the predicted label $\hat{y}$ and the gold label $y$, we compute three statistics: (1) \textbf{OverScore Rate}, the proportion of cases with $\hat{y} > y$; (2) \textbf{UnderScore Rate}, the proportion of cases with $\hat{y} < y$; and (3) \textbf{Mean Score Bias}, defined as the average of $\hat{y} - y$ over the evaluation set. As shown in Table~\ref{tab:bias_analysis}, the Phase~II model exhibits a pronounced upward scoring tendency, with an OverScore Rate of 0.5023, an UnderScore Rate of 0.0683, and a Mean Score Bias of 0.5602. This indicates that after supervised reasoning alignment, the model still tends to assign overly high relevance labels, especially on ambiguous or hard-boundary cases.

After applying DGPO in Phase~III, the error profile becomes substantially more balanced. The OverScore Rate drops from 0.5023 to 0.3356, and the Mean Score Bias decreases from 0.5602 to 0.2378, showing that Phase~III effectively suppresses systematic overestimation of relevance. Meanwhile, the UnderScore Rate increases from 0.0683 to 0.1423, suggesting that the model becomes more conservative in a subset of cases. Nevertheless, this trade-off is beneficial overall, as it reduces the strong one-sided upward bias of the Phase~II model and leads to better-calibrated final relevance decisions. These results suggest that the gain of DGPO comes not only from generic preference learning, but also from its ability to correct systematic decision-level deviations that remain after supervised reasoning alignment.

\section{Discussion}

Our results suggest that knowledge-intensive relevance modeling benefits from being decomposed into three complementary stages: query-centered grounding, reasoning alignment, and decision calibration. Rather than treating relevance judgment as a direct label prediction problem, \textsc{RAG-Match} improves it progressively by first strengthening semantic grounding, then aligning the model with structured relevance reasoning, and finally correcting residual decision-level deviations. This staged design helps explain why the full framework consistently improves ranking performance on challenging search scenarios that require background knowledge, implicit intent understanding, and fine-grained relevance discrimination.

A first important observation concerns the role of Phase~I. We do not interpret Phase~I as a standalone ranking stage, since its objective is not to directly optimize the final ranking metrics. Instead, its value lies in providing a knowledge-augmented initialization for downstream reasoning alignment. More specifically, the query-to-synthesized-document objective encourages the model to build a query-centered semantic scaffold that captures implicit intent, relevant background knowledge, and the semantic content structure that a relevant document is expected to satisfy. Compared with direct label supervision, this generative objective offers denser and more structured training signals, because it teaches the model not only whether a document is relevant, but also what information a query should evoke.

This perspective helps explain the empirical pattern observed in our experiments. Adding Phase~I before Phase~II leads to further improvements in NDCG, while the nPNR results remain largely comparable to those of Phase~II alone. This suggests that Phase~I mainly improves overall ranking quality by strengthening semantic grounding and enriching query-centered background knowledge before reasoning alignment. In contrast, pairwise ranking consistency appears to be more directly shaped by the explicit hierarchical reasoning supervision introduced in Phase~II, which may explain why Phase~I brings only limited additional gains on nPNR. In this sense, the main contribution of Phase~I is not direct decision-boundary learning, but the construction of a more informative semantic starting point for downstream discriminative relevance modeling.

A second observation concerns the role and limitation of the teacher model in Phase~II. Our manual inspection shows that GPT-5.2 produces highly reliable reasoning trajectories, achieving a CoT accuracy of 93.3\%, while its relevance label accuracy is lower at 76.6\%. This suggests that the teacher is particularly effective for structured reasoning supervision, but its final label decisions still contain non-negligible noise. Consistent with this observation, we find that after Phase~II training, the student model reaches a CoT accuracy that is largely comparable to that of GPT-5.2, indicating that the reasoning capability is effectively distilled. However, case analysis reveals that some examples contain correct CoT explanations but incorrect final labels. This mismatch implies that reasoning quality and decision quality are not always perfectly aligned, and also reveals a potential source of error propagation in supervised reasoning alignment.

This issue becomes clearer when we examine the error distribution of the Phase~II model. Our additional analysis shows that Phase~II exhibits a pronounced upward scoring tendency in final relevance prediction, with an OverScore Rate of 0.5023, an UnderScore Rate of 0.0683, and a Mean Score Bias of 0.5602. In other words, the model is much more likely to overestimate relevance than to underestimate it, especially on ambiguous or hard-boundary cases. This provides direct empirical support for the motivation of Phase~III. By introducing discrepancy-guided preference optimization, we encourage the model to prefer more reliable relevance outcomes when competing predictions reveal boundary-level inconsistencies. As a result, Phase~III substantially reduces the OverScore Rate to 0.3356 and the Mean Score Bias to 0.2378, while moderately increasing the UnderScore Rate to 0.1423. Although this makes the model somewhat more conservative in a subset of cases, it effectively mitigates the strong one-sided overestimation tendency inherited from Phase~II and leads to better-calibrated final relevance judgments.

Taken together, these findings suggest that the three phases of \textsc{RAG-Match} play distinct but complementary roles. Phase~I improves semantic grounding and benefits downstream ranking quality; Phase~II contributes the core reasoning capability through hierarchical supervision; and Phase~III serves as a calibration stage that corrects decision-level inconsistencies that cannot be fully resolved by supervised reasoning alignment alone. This division of labor helps explain why the full framework achieves stronger and more stable performance across both ranking quality and relevance decision consistency.

At the same time, our study also highlights several limitations. First, the effectiveness of the framework still depends on the quality of retrieved evidence, and noisy or incomplete retrieval may affect both grounding and reasoning. Second, the current pipeline relies on multi-stage training and teacher-generated supervision, which increases annotation and optimization cost compared with simpler single-stage approaches. Third, although DGPO improves decision calibration, some difficult cases may still contain residual uncertainty, especially when retrieval evidence is incomplete or when the relevance boundary itself is highly ambiguous. These limitations point to the need for more robust retrieval-aware modeling and more efficient alignment strategies in future work.

\section{Conclusion}

In this paper, we proposed \textsc{RAG-Match}, a three-stage framework for knowledge-intensive relevance modeling that integrates knowledge-augmented pretraining, hierarchical reasoning alignment, and discrepancy-guided preference optimization. The central idea is that accurate relevance judgment should be learned progressively: first by strengthening query-centered grounding, then by aligning the model with explicit multi-step relevance reasoning, and finally by calibrating difficult boundary decisions. Experimental results on a real-world search relevance benchmark show that \textsc{RAG-Match} consistently outperforms strong LLM-based baselines across multiple ranking metrics, demonstrating the value of combining knowledge injection, structured reasoning supervision, and preference-based calibration for fine-grained relevance estimation.

Our analysis further shows that the three stages contribute in different ways. Phase~I mainly improves downstream ranking quality by providing stronger semantic grounding for Phase~II, rather than serving as an independently optimized ranking stage. Phase~II effectively distills high-quality reasoning patterns from the teacher model, but still inherits some label-level noise and systematic overestimation in final relevance decisions. Phase~III addresses this issue by reducing the upward scoring tendency of the Phase~II model and producing better-calibrated relevance judgments.

Despite these encouraging results, several limitations remain. The framework is still sensitive to retrieval quality, and noisy or incomplete evidence may weaken downstream grounding and reasoning. In addition, the multi-stage pipeline introduces extra annotation, training, and inference complexity, which may limit efficiency in large-scale or latency-sensitive deployment settings. Future work may explore more robust handling of retrieval noise, for example through evidence filtering, retrieval confidence modeling, or tighter joint optimization between retrieval and relevance estimation. Another important direction is to reduce the latency and complexity of multi-stage relevance modeling, such as by developing more efficient alignment strategies or more unified end-to-end training schemes. We hope this work can provide a useful step toward more accurate, robust, and interpretable relevance modeling for knowledge-intensive search.

\appendix

\section{Prompt Template for Phase\\~II Reasoning Annotation}
\label{app:phase2_prompt}

This appendix provides the prompt template used in Phase~II, i.e., Hierarchical Reasoning Alignment (HRA), to construct teacher-generated reasoning trajectories. As described in Section Method, Phase~II consists of two stages: Knowledge-Infused Grounding (KIG) and Holistic Multi-Dimensional Alignment (HMDA). We use GPT-5.2 as the teacher LLM to generate structured rationales and relevance labels for triplets of the form $\langle q, d, s \rangle$, where $q$ is the query, $d$ is the candidate document, and s denotes the RAG synthesized document.

To improve readability and compatibility with our annotation pipeline, the original prompt was written in Chinese. Below we provide its English version, translated as faithfully as possible while aligning terminology with the main paper.

\subsection{English Prompt Schema Used for Phase~II}

\begin{promptschema}
\small

\noindent\textbf{Role / Instruction.}  
You are an expert in semantic understanding and relevance analysis. Given a \textbf{Query}, \textbf{Retrieved Evidence}, and a \textbf{Candidate Document}, determine the relevance score of the candidate document with respect to the query and explain the scoring logic in detail.

\vspace{4pt}
\noindent\textbf{Important Constraints.}
\begin{itemize}\itemsep2pt
    \item Avoid explicitly mentioning terms such as ``retrieved evidence,'' ``reference knowledge,'' or ``RAG'' in the reasoning whenever possible.
    \item Judge relevance by the extent to which the document satisfies the information need of the query. Retrieved evidence is only used for factual supplementation and verification rather than for literal similarity comparison.
    \item Avoid overly strict entity matching. If the document addresses the same category of core object, the same event chain, or the same thematic set as the query, reasonable partial alignment should still receive partial credit.
    \item If the query does not explicitly specify year, version, region, educational stage, or similar constraints, do not penalize a document merely because other versions may exist, unless there is a clear contradiction or conflict.
    \item Downgrade the score if the document contains off-topic extensions, insufficient coverage, or requires substantial reader-side filtering or correction even when some core information is correct.
\end{itemize}

\vspace{4pt}
\noindent\textbf{Step A: Knowledge-Infused Grounding.}  
Extract from the retrieved evidence the information that is directly relevant to the query and useful for relevance judgment. Summarize it concisely and abstractly rather than copying passages or listing excessive details. The extracted information should include:
\begin{itemize}\itemsep2pt
    \item query intent and task type,
    \item expected answer type,
    \item core object and scope,
    \item minimal answer elements,
    \item key constraints and common pitfalls,
    \item core knowledge points that directly support answering the query.
\end{itemize}

\vspace{4pt}
\noindent\textbf{Step B: Holistic Multi-Dimensional Alignment.}  
Evaluate the candidate document based on the results of Step~A. Focus on semantic alignment and degree of information satisfaction rather than mechanical matching. The evaluation should consider:
\begin{itemize}\itemsep2pt
    \item intent alignment,
    \item entity and key information matching,
    \item knowledge completeness and logical consistency,
    \item proportion of relevant content.
\end{itemize}

\vspace{4pt}
\noindent\textbf{Scoring Criteria.}
\begin{itemize}\itemsep2pt
    \item \textbf{3 (Fully Satisfied):} Covers the minimal answer elements, aligns well with the query intent and object scope, and most of the document is directly useful for answering the query.
    \item \textbf{2 (Mostly Satisfied):} Covers most key elements and answers the core question, but may have minor omissions, slight formulation deviations, or limited extension to closely related secondary objects.
    \item \textbf{1 (Partially Satisfied):} Only touches peripheral aspects, or is related to the topic but misses many key conclusions or essential points.
    \item \textbf{0 (Not Satisfied):} Fundamentally irrelevant to the query, or contains conflicting information that prevents support for the query.
\end{itemize}

\vspace{4pt}
\noindent\textbf{Input Variables.}
\begin{itemize}\itemsep2pt
    \item \textbf{Query}: \texttt{\{query\}}
    \item \textbf{Retrieved Evidence}: \texttt{\{rag\_doc\}}
    \item \textbf{Candidate Document}: \texttt{\{"title": \{title\}, "content": \{content\}\}}
\end{itemize}

\vspace{4pt}
\noindent\textbf{Output Format.}  
Return strict JSON with the following schema:

\vspace{3pt}
{\ttfamily
\noindent\{
"score": "relevance score (0-3)",\\
"cot": "First provide the Step A knowledge extraction in bullet points, then provide the Step B alignment analysis and final scoring rationale in bullet points"\\
\}
}

\end{promptschema}

\subsection{Relation to the Main Framework}

The schema above directly corresponds to the Phase~II design described in the main paper. Specifically, Step~A operationalizes Knowledge-Infused Grounding (KIG) by extracting query intent, expected answer type, core object range, minimal answer elements, key constraints, and core knowledge points from the retrieved evidence. Step~B operationalizes Holistic Multi-Dimensional Alignment (HMDA) by evaluating the candidate document with respect to intent alignment, entity-level matching, knowledge completeness, and the proportion of relevant content. The resulting teacher-generated outputs are then used to construct the reasoning-annotated dataset for supervised fine-tuning in Phase~II.

\bibliographystyle{IEEEtran}
\bibliography{reference}

\end{document}